\documentstyle{l-aa}     
\newcommand{\be}{\begin{equation}}
\newcommand{\ee}{\end{equation}}
\newcommand{\bd}{\begin{displaymath}}
\newcommand{\ed}{\end{displaymath}}
\newcommand{\gapprox}{\;\rlap{\lower 2.5pt                              
             \hbox{$\sim$}}\raise 1.5pt\hbox{$>$}\;}                    
\newcommand{\lapprox}{\;\rlap{\lower 2.5pt                              
             \hbox{$\sim$}}\raise 1.5pt\hbox{$<$}\;}

\newcommand {\lii} {Li\,{\sc i}\,$\lambda$6708~}  
\newcommand {\li} {Li\,{\sc i}~}

\newcommand {\kms} {km~s$^{\rm -1}$}
\newcommand{\vsini}{$v$sin$i$ }

\newcommand {\ha} {H$\alpha$~}
\newcommand {\msun} {M$_{\odot}$~}
\newcommand {\tef} {T$_{\rm eff}$~}
\def\x{$\pm$}  
\begin{document}
   \thesaurus{06         
              (08.01.1;  
                08.05.3; 
                08.09.3; 
                08.12.1  
                08.16.5  
                08.18.1)  
}
   \title{\ha emission fluxes and lithium abundances of low mass 
stars in the young open cluster IC~4665\thanks
{Based on observations made with the  Isaac Newton telescope,
operated on the island of La~Palma by the Royal Greenwich Observatory in the
Spanish Observatorio del Roque de los Muchachos of the Instituto de
Astrof\'\i sica de Canarias, and with the 3.5 m telescope of the 
German-Spanish Calar Alto Observatory.}}
  
 \author{E.L. Mart\'{\i}n \inst{1} \and  D. Montes \inst{2}  }
\institute{Instituto de Astrof\'{\i}sica de Canarias, E-38200 La Laguna, 
Tenerife, Spain \\
\and Departamento de Astrof\'{\i}sica, Facultad de F\'{\i}sicas,
 Universidad Complutense de Madrid, E-28040 Madrid, Spain}

 \offprints{E.L. Mart\'{\i}n, e-mail: ege@iac.es}

   

  \maketitle



   \begin{abstract}

As part of a long term effort to understand pre-main sequence Li burning, 
we have obtained high resolution spectroscopic observations of 14 
late type stars (G0--M1) in the young open cluster IC~4665. 
Most of the stars have \ha filled-in and \li absorption, as expected 
for their young age. 
From the equivalent widths of \ha emission excess 
(obtained using the spectral subtraction 
technique) and the \lii feature, we have derived \ha emission fluxes and 
photospheric Li abundances. 
The mean Li abundance of IC~4665 
solar-type stars is log N(Li)=3.1; the same 
as in other young clusters ($\alpha$~Per, Pleiades) and T Tauri stars. 
Our results support the conclusions from previous works 
that PMS Li depletion is very small for masses $\sim$ 1 \msun .
Among the IC 4665 late-G and early K-type stars, 
there is a spread in Li abundances 
of about one order of magnitude. The Li-poor IC~4665 members have low 
\ha excess and vsin{\it i}$\le$10. Hence, the Li-activity-rotation connection 
which has been clearly established in the Pleiades also seems to hold 
in IC 4665. 
One M-type IC~4665 star that we have 
observed does not show Li, implying a very efficient Li depletion as 
observed in $\alpha$~Per stars of the same spectral type.  

The level of chromospheric activity and Li depletion among  
the low mass stars of IC 4665 is similar to that in the Pleiades.  
In fact, we note that the Li abundance distributions in several young clusters 
 ($\alpha$~Per, Pleiades, IC~2391, IC~4665)  and in post T Tauri stars 
are strikingly similar. 
This result suggests that        
\ha emission and Li abundance not well correlated with   
age for low mass stars between 20 and 100 Myr old.  
We argue that a finer age indicator, the ``LL-clock", 
would be the luminosity at which the transition between efficient 
Li depletion and preservation takes place for fully convective objects.  
The LL-clock  could allow in the near future to derive the relative 
ages of young open clusters, and clarify the study of PMS evolution 
of cool stars.


\end{abstract}

\section{Introduction}

This paper is the continuation of an ongoing project aimed at understanding  
pre-main sequence (PMS) Li burning. In previous works we have studied 
Li in different types of PMS stars: 
Mart\'\i n et al. (1992) obtained Li abundances 
for 5 post T Tauri secondaries of early-type stars.  
Mart\'\i n \& Rebolo (1993) studied the PMS secondary of the eclipsing 
binary EK~Cep and derived its Li abundance. 
Mart\'\i n et al. (1994) presented Li abundances for 53 ``weak'' 
T Tauri stars (WTTS) in Taurus. 
They found a narrow peak in the Li distribution centered at 
log N(Li)$_{\rm NLTE}$=3.1, which is the initial Li content 
of newly formed stars. They also found that PMS Li depletion is 
a strong function of mass and luminosity. 
Garc\'\i a L\'opez et al. (1994) derived Li 
abundances for 24 Pleiades stars and confirmed the 
presence of a Li-rotation connection (see also Soderblom et al. 1993) 
among the late-G and early K-type stars, but not among 
the late-K and early-M stars. Such connection   
has been studied theoretically by Mart\'\i n \& Claret (1996), 
who showed that rotation lowers the temperature at the base of the 
convection region and hence reduces the efficiency of PMS Li depletion. 
Finally, Zapatero-Osorio et al. (1996) performed a search for Li 
in very low mass (M3--M6) $\alpha$~Per stars with no positive Li detection. 

The IC~4665 open cluster offers the possibility of testing our current 
ideas about PMS Li burning because it may have an age of only $\sim$35 Myr 
(Mermilliod 1981, but see the discussion in Section 4.2).  
Its coordinates (17h40m, +5$^o$) make it a suitable target from 
northern hemisphere observatories. Its distance ($\sim$350pc) is almost 
a factor of 3 farther than the Pleiades, and thus the cluster members 
are relatively faint. Nevertheless, a list of reliable 
low mass members with V-magnitudes in the range 12--14 has 
been released from proper motion and radial velocity studies (Prosser \& 
Giampapa 1994). We selected for spectroscopic observations the 
13 bona fide IC~4665 
members with colors (B-V)$>$0.75, i.e. spectral type G0 and later, given 
by Prosser \& Giampapa. We added to this sample one early M-type 
photometric member from Prosser (1993). 

Our spectra contain not only the \lii feature, but also \ha . 
The \ha emission excess is a fairly 
good indicator of chromospheric activity 
(Pasquini \& Pallavicini 1991, Montes et al. 1995a),  
which in turn is connected to rotation via the dynamo mechanism. 
In the Pleiades cluster the fast rotators present both high Li abundance and 
high level of chromospheric activity (Soderblom et al. 1993, Garc\'\i a 
L\'opez et al. 1994). Hence, there is a Li-activity connection induced 
by rotation. We have tested if such connection also holds for 
the IC~4665 stars.

\section{Spectroscopic observations}

All the programme stars, except P75, were observed at the 2.5m Isaac Newton 
telescope (INT) with the Intermediate Dispersion Spectrograph (IDS) 
during a four night run in August 1995. We used the 235mm camera, AgRed 
collimator, gratings R1200Y and H1800V and a TEK 1124x1124 CCD detector. 
For each star we give in Table~1 the Prosser (1993) number and V-magnitude, 
date of the observation, 
dispersion (resolution $\sim$ 2 pixels), and exposure time. 
One of the stars was observed with the 3.5m telescope at Calar Alto 
observatory, using the TWIN spectrograph, T06 grating in the red arm and 
a TEK chip. 

\begin{table*}
\caption[]{Observing log}
\begin{tabular}{llllll}
\hline
\hline
Star & V & Date & Tel. & Disp. & Texp \\
     &  &       &     & (\AA ~ pix$^{-1}$) & (s) \\
\hline
\hline
P12 & 12.72 & Aug 6, 1995 & INT & 0.53 & 1200 \\
P27 & 12.65 & Aug 6-9 1995 & INT & 0.53-0.84 & 3300$^*$ \\
P60 & 13.43 & Aug 9, 1995 & INT & 0.53 & 900 \\
P71 & 13.68 & Aug 6, 1995 & INT & 0.53 & 1600 \\
P75 & 13.70 & Jul 17, 1995 & Calar & 0.91 & 1000 \\
P94 & 14.26 & Aug 8, 1995 & INT & 0.84 & 1200 \\
P100 & 14.37 & Aug 7, 1995 & INT & 0.84 & 900 \\
P107 & 12.94 & Aug 8, 1995 & INT & 0.84 & 900 \\
P146 & 14.19 & Aug 6, 1995 & INT & 0.53 & 1800 \\
P150 & 13.08 & Aug 7, 1995 & INT & 0.84 & 600 \\
P155 & 13.52 & Aug 7, 1995 & INT & 0.84 & 900 \\
P165 & 13.40 & Aug 8, 1995 & INT & 0.84 & 900 \\
P166 & 13.79 & Aug 8, 1995 & INT & 0.84 & 900 \\
P309 & 16.80 & Aug 7-8, 1995 & INT & 0.84 & 3000 \\
\hline         
\hline
\end{tabular}
\vskip2truept
Note: the exposure time for P27 is the sum of 4 individual exposures. 
This star is a spectroscopic binary. 
\end{table*}

Each spectrum was reduced by a standard procedure using 
IRAF\footnote{IRAF is distributed by the National Optical Observatory, which is 
operated by the Association of Universities for Research in Astronomy, Inc., 
under contract with the National Science Foundation.}, which included 
debias, flat field, optimal extraction and wavelength calibration 
using CuNe arc lamps. Figure 1 displays the full range of the final spectra. 

\begin{figure}
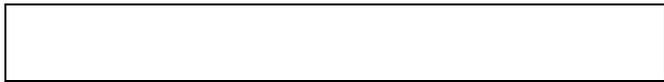
 
\picplace {1 cm} 
\caption[]{Spectra of 14 late-type stars in the open cluster IC~4665. 
The positions of \ha and \lii are marked.} 
\end {figure}

\section{Analysis}

The Li abundances for our programme stars have been derived from 
the \lii resonance line. We have measured the equivalent widths of the Li  
feature by direct integration and gaussian profile fitting. Both authors 
of this paper measured the equivalent width independently and we took 
the mean as the final value. Typically, the standard deviation 
of different measurements was $\sim$ 15 m\AA , which is the value that 
we took as error bar. 
We used NLTE curves of growth for the \li line from Mart\'\i n et al. 
(1994). The validity of these curves has been confirmed by the more 
recent work of Pavlenko et al. (1995). 
Effective temperatures for the IC~4665 stars were derived by two different 
methods: a calibration of B-V colour vs. \tef (Arribas \& Martinez 
Roger 1988), and a calibration of spectral type vs. \tef 
(de Jager \& Nieuwenhuijzen 1987). The B-V colours were taken from 
Prosser (1993) and they were corrected 
by the average cluster colour excess of E(B-V)=0.18. 
The spectral types were derived from comparison of our spectra with a 
battery of standards observed with the IDS (Montes et al. 1995b).  
We estimate that our spectral types are accurate to about half a spectral 
subclass. The temperature of P309 was estimated from the spectral type only,  
because the B-V calibration of Arribas \& Martinez Roger did not 
include M-type stars. 

 The differences in the \tef obtained from (B-V) and spectral type 
ranged from 30 to 310~K, with an average of 147~K. 
For the star P12 we found an anomalous large discrepancy 
between the colour and the spectral type. This star was classified 
by Prosser \& Giampapa (1994) as G0V and they also gave a \vsini =10\kms . 
In contrast, we find a spectral type of K0V and a \vsini =70\kms . 
Our high \vsini is consistent with the short rotational period 
found by Allain et al. (1996) of 0.60 days. We believe that there 
is a problem with the photometry and spectroscopy given by Prosser (1993) 
and Prosser \& Giampapa (1994) for this star. Hence, we chose to ignore 
the photometry and used only our spectral type for deriving its \tef .

 Taking into account the maximum uncertainties in \tef (\x 150 K) 
and \li equivalent widths (\x 15 m\AA ), they lead to an  
error bar in the Li abundances of $\pm$ 
0.3 dex. This is similar to previous works on Li abundances in 
other clusters (e.g. Garc\'\i a L\'opez et al. 1994). 
In Table~2 we give the NLTE Li abundances of our programme stars in the 
customary scale of log N(H)=12.

\begin{table*}
\caption[]{\ha fluxes and Li abundances of IC~4665 stars}
\begin{tabular}{lllllllll}
\hline
\hline
Star & B-V & SpT. & \tef & \vsini & EW \ha & log F(\ha) & EW \li 
& log N(Li) \\
     &  &  & (K) & (\kms ) & (\AA )  & (erg cm$^{-2}$ s$^{-1}$ \AA $^{-1}$)& 
(m\AA ) &  \x 0.3 dex \\
\hline
\hline
P12 & 0.77 & K0 & 5150$^*$ &   70 & 0.475 & 6.36 & 270 & 2.9 \\
P27 & 0.77 & G0 & 5900 &   33 & 0.076 & 5.78 & 140 & 3.1 \\
P60 & 0.88 & G7 & 5440 &   13 & 0.261 & 6.20 & 215 & 3.0 \\
P71 & 0.92 & G8 & 5340 &   17 & 0.602 & 6.54 & 260 & 3.1 \\
P75 & 0.89 & G9 & 5345 &  16 & 0.592 & 6.51 & 320 & 3.3 \\
P94 & 1.01 & K0 & 5135 &   10 & 0.198 & 5.99 &  90 & 2.1 \\
P100 & 1.06 & K0 & 5070 &  21 & 1.184 & 6.72 & 277 & 2.8 \\
P107 & 0.84 & G1 & 5745 & 27 & 0.044 & 5.51 & 156 & 3.0 \\
P146 & 1.12 & K0 & 5000 & $<$10 & $\sim$0.0 &  & $<$40 & $<$1.6 \\
P150 & 0.86 & G7 & 5470 & 25 & 0.721 & 6.65 & 235 & 3.1 \\
P155 & 0.91 & G7.5 & 5365 & 17 & 0.415 & 6.39 & 264 & 3.2 \\
P165 & 0.90 & G9 & 5330 & 40 & 0.794 & 6.63 & 253 & 3.1 \\
P166 & 0.97 & K0 & 5190 & $<$10 & 0.328 & 6.20 & $<$35 & $<$1.4 \\
P309 & 1.62 & M1 & 3660$^*$ & $<$20 & 1.439 & 6.00 & $<$170 & $<$0.5 \\
\hline         
\hline
\end{tabular}
\vskip2truept
Note: $^*$ effective temperature derived only from the spectral type 
(see text for remarks on P12).  
\end{table*} 

The \ha emission excess was measured using the spectral subtraction 
technique. Reference spectra were constructed for all the IC 4665 stars 
using main-sequence standards of the same spectral type, which were 
rotationally broadened and doppler shifted to give as good match as possible. 
We subtracted the reference spectra from the IC~4665 spectra and 
obtained the chromospheric contribution of \ha . For more details 
on this technique see Montes et al. (1995a). 
In Figure~2 we display one example (P12) of the spectral subtraction method. 
In the subtracted spectra  \ha shows up in emission and \li  
in absorption, whereas the other photospheric lines are canceled.   
The equivalent width of \ha in emission were converted to flux 
using: a calibration of B-V vs. surface flux in the \ha region 
(Hall 1996); and 
 a calibration of V-R vs. surface flux in the \ha region (Pasquini \& 
Pallavicini 1991). We obtained the B-V and V-R colours from our spectral 
types and from dereddening the B,V photometry of Prosser (1993). 
The results obtained from both colours agreed quite well,
and the final values adopted are given in Table~2. 
We note that for P27 the radial velocity changed from night to night and 
we also noticed variations in the \ha emission from logF(\ha)=5.63 to 
log F(\ha)=5.86. 
 In all our spectra of P27 the \li equivalent widths 
were similar except in that with the largest \ha emission, which 
presented a shallower \li feature (equivalent width 110 m\AA ).  
We suggest that P27 deserves monitoring of radial velocity, \ha and Li 
variations.  

\begin{figure}
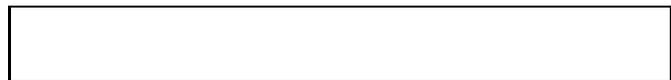
 
\picplace {1 cm} 
\caption[]{One example of the spectral subtraction technique. 
The upper spectrum is 
the subtracted spectrum. The lower spectra are the observed IC 4665 star
(solid line) and superimposed the reference spectrum (dashed line).
In the small window we show a zoom of the H$\alpha$ region.
The upper spectra are the subtracted spectrum (dotted line) and 
a Gaussian profile fit (dotted-dashed line). 
The lower spectra are the IC 4665 star (solid line)
and the standard (dashed line).} 
\end {figure}

\section{Discussion}

\subsection{Chromospheric activity and Li depletion in IC 4665}

We have found a large spread in \ha chromospheric fluxes and 
photospheric Li abundances 
among the IC~4665 stars. Figure~3 displays the temperature dependance 
of \ha and Li. They show different behaviours: \ha emission tends to increase  
towards cooler \tef, whereas Li decreases. Both effects can be 
qualitatively explained 
by the increase in the depth of the convection region towards cooler 
stars. Note that the spread both in \ha and Li appears to increase 
for \tef $<$5500~K. 

\begin{figure}
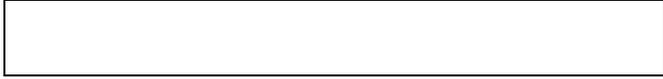
 
\picplace {1 cm} 
\caption[]{Upper panel: 
Excess \ha emission fluxes for the IC~4665 stars versus 
\tef . The star P146 is not plotted because it did not show a measurable 
\ha excess. Lower panel: Li abundances for the IC~4665 stars versus 
\tef .} 
\end {figure}

In order to place our  
results in a wider context we have compared the IC 4665 stars with 
the well-studied Pleiades cluster.  Figure~4 shows our Li abundances 
together with those of Soderblom et al. (1993) and Garc\'\i a L\'opez 
et al. (1994) for the Pleiades. The patterns are very similar.  
Broadly speaking we can distinguish three groups of stars: 

i) early G-type (6000 K--5300 K) stars have quite 
uniform abundances, with a mean 
value of log N(Li)=3.1, which is the same as in WTTS (Mart\'\i n 
et al. 1994) and the interstellar medium. Hence, these stars have retained  
their initial Li abundance. 

ii) late-G and early-K stars 
(5300K--4500K) present a broad range of abundances, which in the 
Pleiades was clearly shown to be related to chromospheric activity and 
rotation (Soderblom et al. 1993; Garc\'\i a L\'opez 
et al. 1994). Such effect could be caused by the effect of rotation 
on the temperature at the base of the convection zone (Mart\'\i n \& 
Claret 1996). 
We only have six stars in this group: P12, P100 and P165 with 
high Li abundances and high \ha excesses, and 
P94, P146 and P166 with low abundances and low excesses. 
In Figure~5, we have plotted the IC~4665 stars from this work, 
and the Pleiades stars from Soderblom et al. (1993) with 
 \tef in the range 5300K--4900K in a Li vs. \ha diagram. 
We converted our \ha fluxes to flux ratios 
(R$_{H\alpha}$=F$_{H\alpha}$/$\sigma$\tef$^4$) in order to compare with 
Soderblom et al. In the stars of both clusters there is 
a similar correlation of Li with \ha excess. The stars P146 and P166 
have very small \ha excesses (we could not measure any excess in the first one 
and it is not included in Figure~5) and we could not detect Li. 
The comparison with the Pleiades 
suggests that such \ha activity and Li are very unusual and it may indicate 
that these stars are in fact not cluster members. 
 However, we hesitate 
to discard P146 and P166 as cluster members because their radial 
velocities match the cluster mean (Prosser \&  Giampapa 1994).   
We note that if these stars 
were non-members, the correlation between Li and \ha 
flux does not disappear, but the specimens with low Li and low \ha excess 
are reduced to only one (P94). Hence, it is important to find 
more low mass IC~4665 members in which we can test better 
the Li-activity connection in this cluster. 

Prosser \& Giampapa (1994) and Allain et al. (1996) obtained vsin{\it i} 
and photometric rotation periods, respectively, for several of our 
programme stars. The three stars with low Li and \ha excess (P94, P146 and 
P166) have vsin{\it i}$\le$10 \kms . Allain et al. could not detect 
any period in P94 and P146. On the other hand the three stars with high 
Li abundance and \ha excess have high vsin{\it i} (see Table~2). 
Interestingly, Allain et al. (1996) were able to derive rotation periods 
for P12 (0.6 days) and P100 (2.27 days). Therefore, there is a real 
difference between the equatorial velocities of these two stars of 
about a factor 5. The \ha emission level of P100 
is a bit higher than that of P12, indicating that 
there is not a good one to one relationship between \ha and 
rotation, but both stars do have higher activity than the slow 
rotators (P94, P146, P166). 
 The Li abundance of P12 is higher than that of 
P100, but the difference is only 0.1 dex which is within our error bars. 
With only these two stars it is not yet possible to test  
the models of Mart\'\i n \& Claret (1996), which predict that fast rotating 
stars deplete less Li than slow rotating stars during their PMS evolution.   
 
\begin{figure}
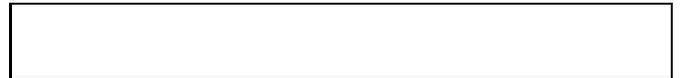
 
\picplace {1 cm} 
\caption[]{Li abundances versus \tef for stars in IC 4665 (diagonal crosses) 
and the Pleiades (empty pentagons). The four upper limits at \tef $<$ 4000 K 
correspond to P309 (dotted trace) 
and 3 $\alpha$~Per stars observed by Garc\'\i a L\'opez 
et al. (1994).} 
\end {figure}

\begin{figure}
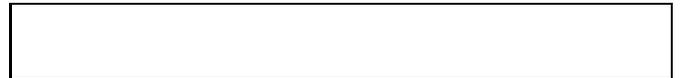
 
\picplace {1 cm} 
\caption[]{Li abundances versus \ha excess emission ratios
 for stars in IC 4665 (diagonal crosses) 
and the Pleiades (empty pentagons) with \tef in the range 
5300 K -- 4900 K.} 
\end {figure}

iii) the late-K and early M stars (4500K--3500K) do not present 
the Li-activity-rotation connection in the $\alpha$Per and 
Pleiades clusters (Garc\'\i a L\'opez 
et al. 1994). We have only observed one M-type star in IC~4665, 
and we could not 
detect Li in it. The inferred upper limit to the Li abundance is shown 
in Figures 3 and 4. Our result is consistent with the high efficiency 
of PMS Li depletion found for the stars of this group. 
The possibility that 
such property could be used for a fine dating of young open clusters 
is discussed in the following section.

\subsection{The age of young clusters and the LL-clock}

Mermilliod (1981) determined an age for IC~4665 of 36 Myr from empirical 
isochrone fitting of the upper main sequence, which is clearly younger 
than the ages he obtained for $\alpha$~Per (51 Myr) and the Pleiades (78 Myr). 
However, some of the stars used by Mermilliod have been shown to be 
binaries in later works, and Prosser (1993) revised the upper main-sequence 
fitting to yield an age of $\sim$ 50--70 Myr.  Prosser concluded that  
IC 4665 is not younger than $\alpha$~Per and it might be as old as the 
Pleiades. 

The distribution of Li abundances in the low mass stars of IC 4665  
and the Pleiades are quite similar (Figure~4). However, this does not 
necessarily imply that both clusters have the same age. The 
 distribution of Li is quite similar 
also in $\alpha$ Per (Balachandran, Lambert \& Stauffer 1998, 
Garc\'\i a L\'opez et al. 1994). 
With respect to \ha activity, in Fig.~6 we compare the IC~4665 stars 
with the Pleiades. The IC 4665 stars present a similar level of 
\ha activity than the Pleiades.

\begin{figure}
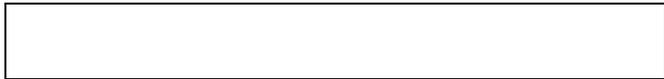
 
\picplace {1 cm} 
\caption[]{\ha excess emission ratios
 for stars in IC 4665 (diagonal crosses) 
and the Pleiades (empty pentagons) versus \tef .} 
\end {figure}

 Stauffer et al. (1989) gave \li equivalent widths for 10 low mass members 
of the IC~2391 cluster, but they did not derive Li abundances. 
This cluster is very interesting because according Mermilliod (1981) it 
belongs to the same age group as IC~4665. We have applied the  
abundance analysis described in Section~3 to the IC~2391 stars. 
The effective temperatures were derived from the Arribas \& Martinez-Roger 
relationship (1988) using the (B-V) colours given by Stauffer et al. (1989), 
dereddened by the mean E(B-V)=0.04 of the cluster.   
The Li abundances obtained are listed in Table~3.   

\begin{table*}
\caption[]{Li abundances for IC~2391 stars}
\begin{tabular}{llllll}
\hline
\hline
Star & B-V & \tef & \vsini & EW \li & log N(Li) \\
     &  & (K) & (\kms ) & (\AA )  &   \x 0.3 dex \\
\hline
\hline
1 & 0.68 & 5685 & 34    & 0.16    & 3.0 \\
2 & 0.57 & 6074 & $<$15 & 0.14    & 3.2 \\
3 & 1.00 & 4793 & 90    & 0.13    & 1.9 \\
4 & 1.39 & 4023 & $<$15 & $<$0.05 & $<$0.4 \\
5 & 1.39 & 4023 & 150   & $<$0.2  & $<$1.2 \\
6 & 0.84 & 5201 & 16    & 0.19    & 2.6 \\
7 & 1.10 & 4568 & $<$15 & $<$0.05 & $<$1.1 \\
8 & 1.25 & 4269 & 18    & 0.09    & 1.0 \\
9 & 1.25 & 4269 & $<$15 & 0.10    & 1.0 \\
10 & 1.44 & 3942 & 95   & $<$0.1 & $<$0.7 \\
\hline         
\hline
\end{tabular}
\vskip2truept
Note: the data was taken from Stauffer et al. (1989). 
\end{table*}

The Li abundances of the IC~2391 stars fall within the range of 
abundances of the IC~4665 and Pleiades stars (Fig.~4).    
We can also compare with 
the post T Tauri stars (PTTS) studied by Mart\'\i n et al. (1992) 
and Mart\'\i n (1993). 
The estimated ages of PTTS are 20--50 Myr, and the Li abundances for 9 PTTS   
are in the range log N(Li)=3.3 -- 1.8, their \tef 
ranging from 5900~K to 4400~K. The degree of Li depletion among the 
PTTS is similar than among the IC~2391, IC~4665, $\alpha$ Per and 
the Pleiades stars, 
indicating that either the PTTS and the 4 open clusters  
share similar age distributions, or Li is not sensitive for ages in the range  $\sim$~20--100~Myr. 
The first interpretation would imply  a rather large age spread within 
the low mass population of open clusters. On the other hand,  
the strong correlation of Li with luminosity found by  Mart\'\i n et al. (1994) 
in WTTS, indicates that PMS Li burning is very efficient during 
a short timescale. It is conceivable that the observed pattern of Li 
abundances in PTTS and young clusters is formed quite rapidly through 
PMS convective mixing (timescale of a few Myr) at an age of  
$\sim$ 20 Myr.  Theoretical models are qualitatively consistent with 
this scenario, because they predict a sudden drop in Li abundance 
when the bottom of the convection region reaches high enough temperature 
for Li burning, but as the star approaches the ZAMS, convection 
becomes shallower and Li depletion should slow down. However, 
it has been noted that theoretical models do not give yet a quantitative 
good description of the observed Li abundances among PMS stars 
(Mart\'\i n et al. 1994, Garc\'\i a L\'opez et al. 1994). 

In Fig.~4, 
it can be seen that below \tef 4000 K, Li has not been detected in any 
star of the $\alpha$~Per and IC 4665 clusters. The upper limits imply 
very strong Li depletions. However, as we move to lower 
masses, the internal temperatures diminish and we should reach 
a point where they are not high enough for Li burning. Hence, we should 
see Li again in the very low mass cluster members. This effect has 
been proposed as a test for distinguishing between stars and brown dwarfs 
(Rebolo et al. 1992; Magazz\`u et al. 1993). 
Very recently, Li has been detected in the coolest known members of the 
Pleiades  (Basri et al. 1996, Rebolo et al. 1996). 
For the age of the Pleiades (70-120 Myr), theoretical models predict that only 
brown dwarfs preserve lithium (e.g. Magazz\`u et al. 1993), but for 
younger ages the very low mass stars also start to preserve it because they  
require a long time to contract. Thus, we expect that in clusters 
younger than the Pleiades, the 
reappearence of Li should be shifted to higher  masses. 
Such an effect provides a precision clock for dating clusters. 
 D'Antona \& Mazzitelli (1994) suggested on the basis of their 
theoretical isochrones of Li depletion that low luminosity stars 
could be used for dating open clusters. They also noted that 
the Li depletion is quite sensitive to input physics of the models which 
have considerable uncertainties (opacities, convection). Hence, 
it may be difficult to use Li to assign absolute ages to open clusters, 
but relative ages will probably be safe.     

What we call in this paper ``LL-clock" stands for Lithium-Luminosity clock.  
It is based on the high efficiency of Li depletion for fully 
convective low luminosity objects. 
The age of a young cluster will be given by the most 
massive objects 
among the very low mass members in which Li is seen to re-emerge, 
after having been efficiently depleted by higher mass cluster members. 
In the Pleiades the Li destruction--preservation 
borderline has been found at luminosity around log~L=-2.9~L$\odot$ 
(Basri et al. 1996) according to the Li detection in the object PPl~15. 
The traditional method for dating open clusters, i.e.   
upper main-sequence turn off fitting, says that 
$\alpha$Per is younger by $\sim$20~Myr than the Pleiades (Mermilliod 1981). 
If PPl~15 were 20~Myr younger, it luminosity would be about 0.12~dex 
larger (D'Antona \& Mazzitelli 1994), and it should be slightly hotter 
and have higher Li abundance. 
Zapatero-Osorio et al. (1996) have reported a negative Li detection in 
AP~J0323+4853, which has a luminosity around log~L=-2.6~L$\odot$. 
This result does not rule out that  $\alpha$Per is younger than 
the Pleiades, but it shows how close the observations are to telling 
us very interesting things about the relative ages of the young clusters. 
In IC~4665, it will be a little more difficult to investigate the 
very low-luminosity members, simply because of its larger distance, but 
if it is indeed younger than the Pleiades, we expect that Li should re-appear 
at higher luminosities.

\section{Conclusions}

We have obtained high resolution spectra of 14 
cool stars (G0--M1) in the open IC 4665 cluster. Excess \ha emission 
and Li abundances have been derived from the data. We find a large 
spread in both parameters. The dependance of \ha and Li with 
\tef in IC 4665 seems similar than in the Pleiades.  Two IC 4665 stars 
(P146 and P166) with lower \ha emission and Li abundance than Pleiades stars 
of the same \tef are suspects of not being cluster members, although 
their radial velocities do indicate membership. 
The Li abundances of IC 4665 stars support previous conclusions on 
PMS Li burning (Mart\'\i n et al. 1994; Garc\'\i a L\'opez et al. 1994): 
solar type stars experience little PMS Li depletion; late-G and early-K 
stars present a large spread of Li abundances (about 1 dex), which 
is related to activity and rotation in the sense that fast rotators 
have higher Li abundance and chromospheric emission; late-K and early-M 
stars experience very efficient Li depletion, which is 
larger than a factor of 100 for \tef $\le$ 4000~K. 

 The distribution of Li abundances 
in the low mass stars of several open clusters 
(IC~2391, IC 4665, $\alpha$~Per, Pleiades), and in post~T~Tauri stars, 
are rather similar. There can be two reasons for that: (a) the stars 
in different clusters and the PTTS  
span the same range of ages, and (b) Li and \ha ~are not  
precise age indicators between $\sim$ 20 and 100 Myr. We argue that 
relative ages of young clusters may be 
obtained in the near future using the ``LL-clock". This would certainly 
help to clarify the evolution of Li and chromospheric activity during late PMS 
evolution.  

\begin{acknowledgements} 

It is a pleasure to thank Artemio Herrero and the staff of the Isaac Newton 
telescope for helping us to carry out the observations. 
DM acknowledges the support by the Spanish Direcci\'{o}n General de 
Investigaci\'{o}n Cient\'{\i}fica y T\'{e}cnica (DGICYT) under grant PB94-0263.

\end{acknowledgements}


\newpage
\begin{thebibliography}{}



\bibitem{} Allain, S., Bouvier, J., Prosser, C., Marschall, L.A., Laaksonen, B.D. 1996, A\&A, 305, 498

\bibitem{} Arribas, S., Martinez Roger, C. 1988, A\&A, 206, 63

\bibitem{} Balachandran, S., Lambert, D.L., Stauffer, J.R. 1988, ApJ, 333, 267 

\bibitem{} Basri, G., Marcy, G.W., Graham, J.R. 1996, ApJ, 458, 600

\bibitem{} D'Antona, F., Mazzitelli, I. 1994, ApJS, 90, 467

\bibitem{} Garc\'\i a L\'opez, R.J., Rebolo, R., Mart\'\i n, E.L. 1994, 
A\&A, 282, 518 

\bibitem{} Hall, J.C. 1996, PASP, 108, 313

\bibitem{} de Jager, C., Nieuwenhuijzen, H. 1987, A\&A, 177, 217 

\bibitem{} Magazz\`u, A., Mart\'\i n, E.L., Rebolo, R. 1993, ApJ, 404, L17

\bibitem{} Mart\'\i n, E.L. 1993, PhD Thesis, Universidad de La Laguna 

\bibitem{} Mart\'\i n, E.L., Rebolo, R. 1993, A\&A, 274, 274  

\bibitem{} Mart\'\i n, E.L., Claret, A. 1996, A\&A, 306, 408  

\bibitem{} Mart\'\i n, E.L., Magazz\`u, A., Rebolo, R. 1992, A\&A, 257, 186 

\bibitem{} Mart\'\i n, E.L., Rebolo, R., Magazz\`u, A., Pavlenko, Ya.V. 
1994, A\&A, 282, 503  

\bibitem{} Mermilliod, J.C. 1981, A\&A, 97, 235

\bibitem{} Montes D., Fern\'{a}ndez-Figueroa M.J., De Castro E.,
   Cornide M. 1995a, A\&A, 294, 165

\bibitem{} Montes D., Fern\'{a}ndez-Figueroa M.J., De Castro E.,
   Cornide M. 1995b, A\&AS, 109, 135

\bibitem{} Pasquini, L., Pallavicini, R. 1991, A\&A, 251, 199

\bibitem{} Pavlenko, Y.V., Rebolo, R., Mart\'\i n, E.L., 
Garc\'\i a L\'opez, R.J. 1995, A\&A, 303, 807

\bibitem{} Prosser, C.F. 1993, AJ, 105, 1441

\bibitem{} Prosser, C.F., Giampapa, M.S. 1994, AJ, 108, 964

\bibitem{} Rebolo, R., Mart\'\i n, E.L., Magazz\`u, A. 1992, ApJ, 389, L83

\bibitem{} Rebolo, R. et al. 1996, in preparation

\bibitem{} Soderblom, D.R., Jones, B.F., Balachandran, S., Stauffer, 
J.R., Duncan, D., 
Fedele, S.B., Hudon, J.D. 1993, AJ, 106, 1059 

\bibitem{} Stauffer, J.R., Hartmann, L.W., Jones, B.F., McNamara, B.R. 
1989, ApJ, 342, 285 

\bibitem{} Zapatero-Osorio, M.R., Rebolo, R., Mart\'\i n, E.L., 
Garc\'\i a L\'opez, R.J. 1996, A\&A, 305, 519




\end{thebibliography}
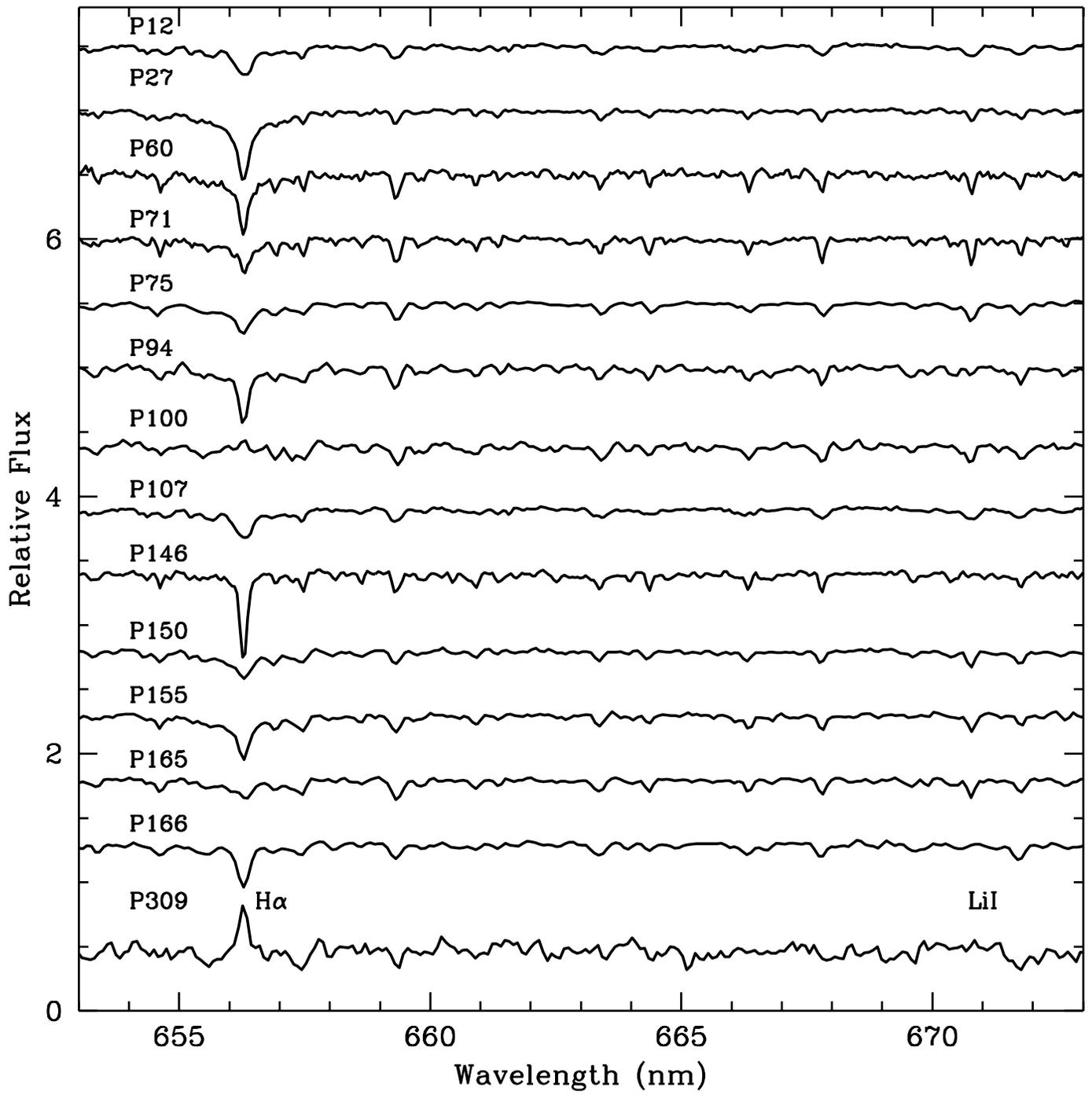
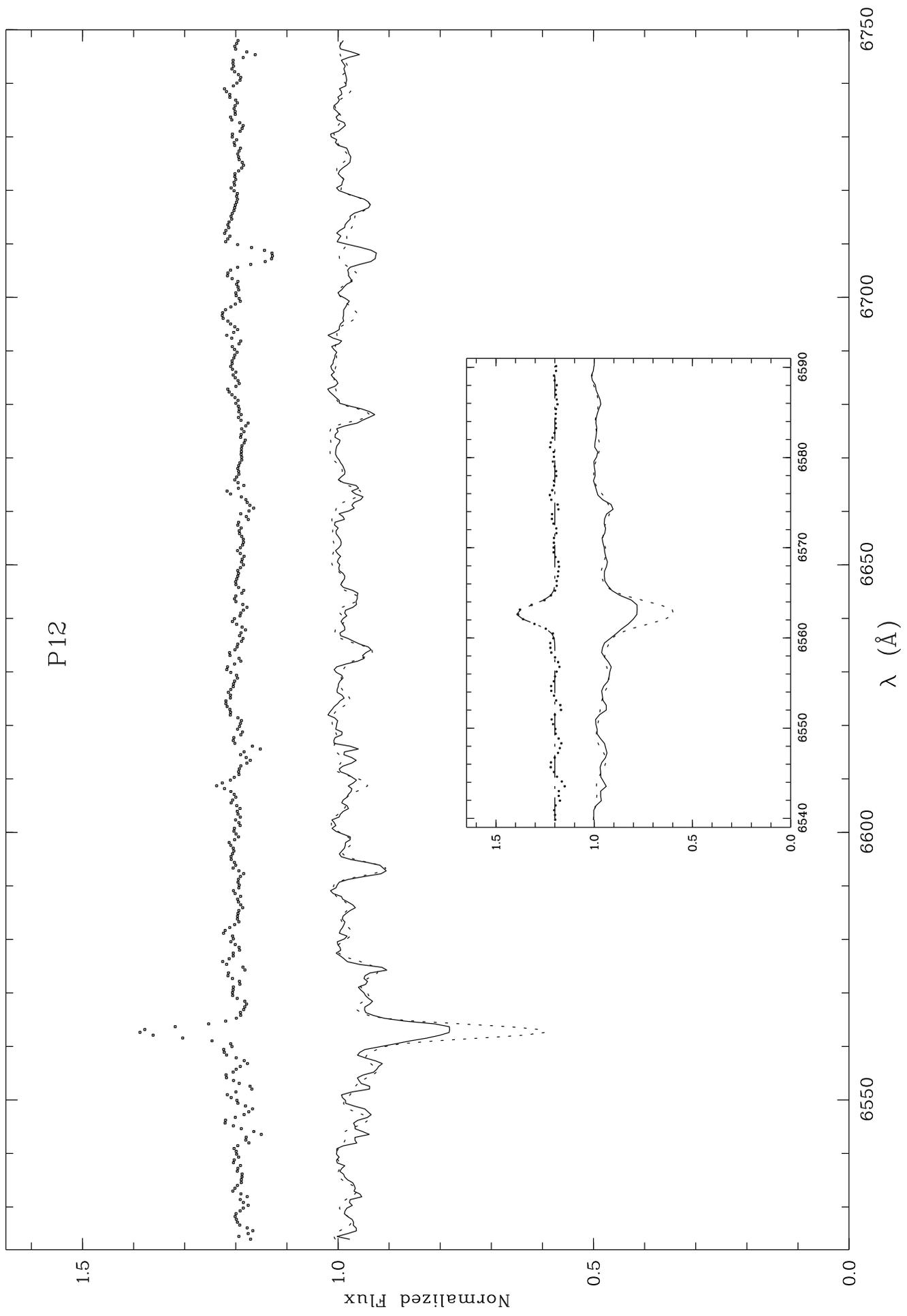
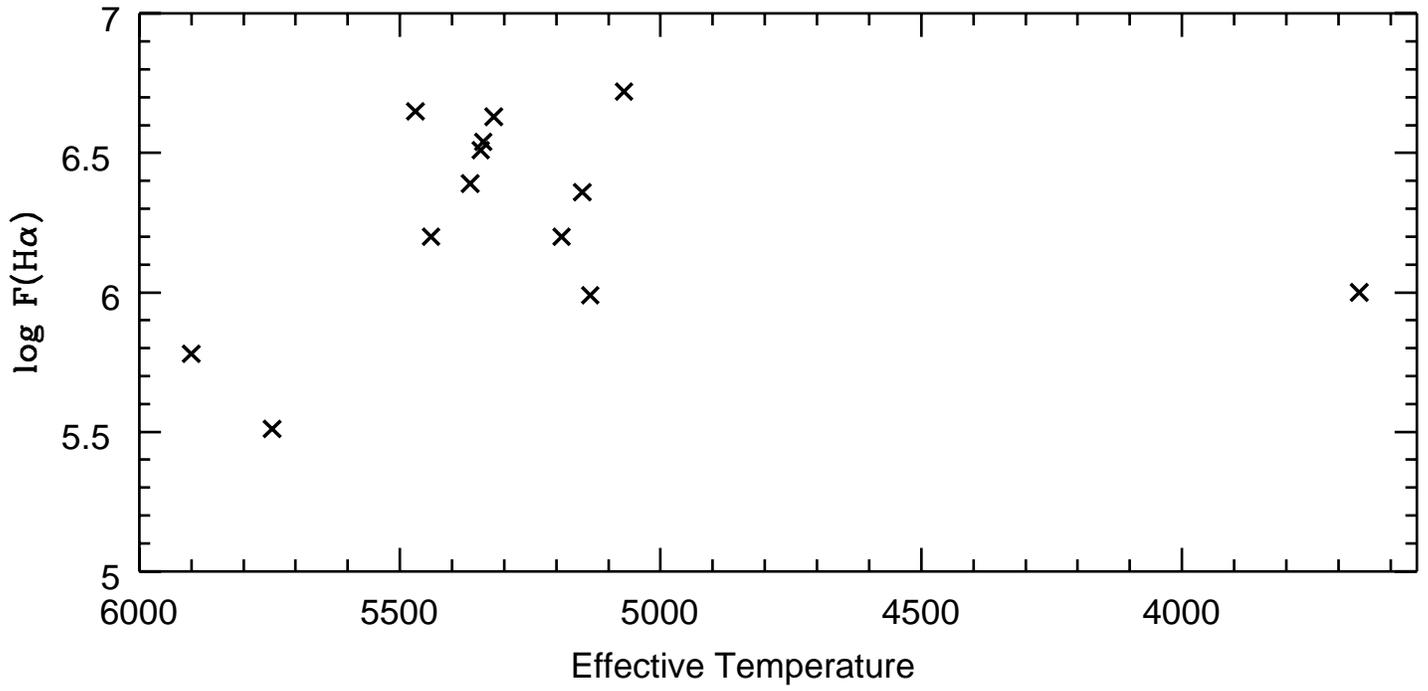
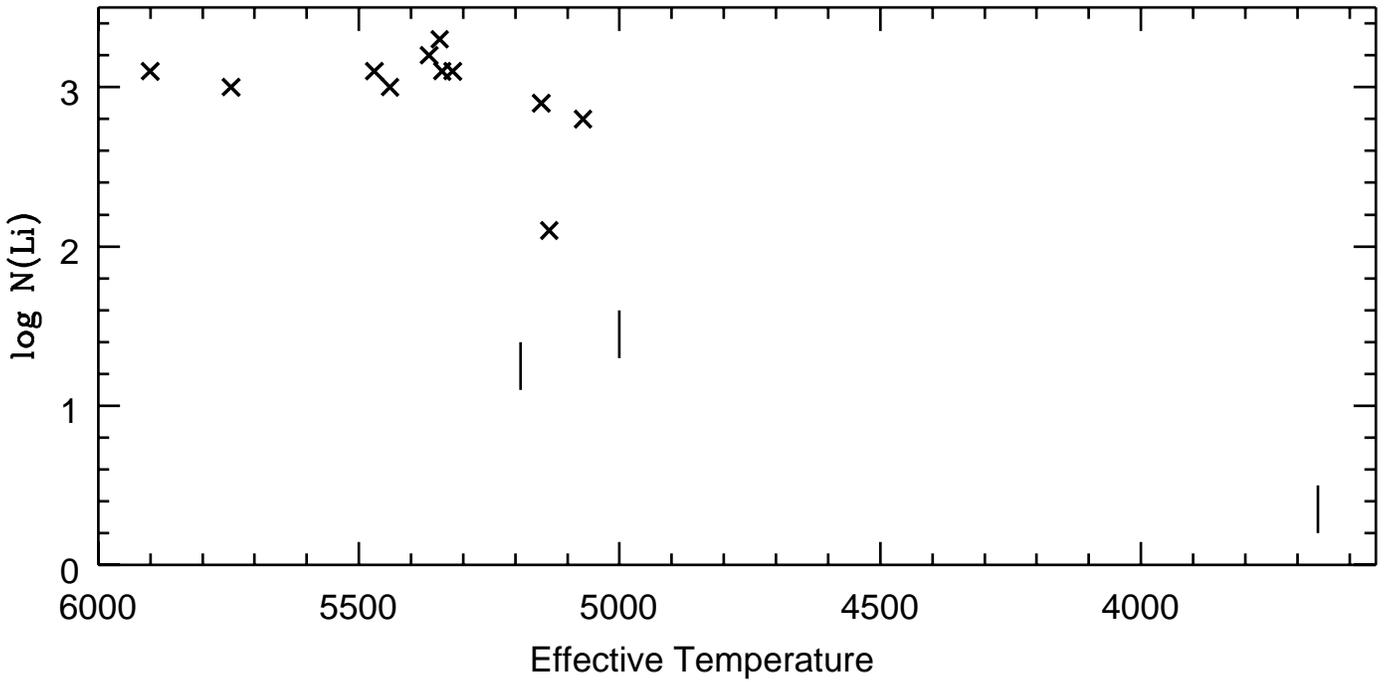
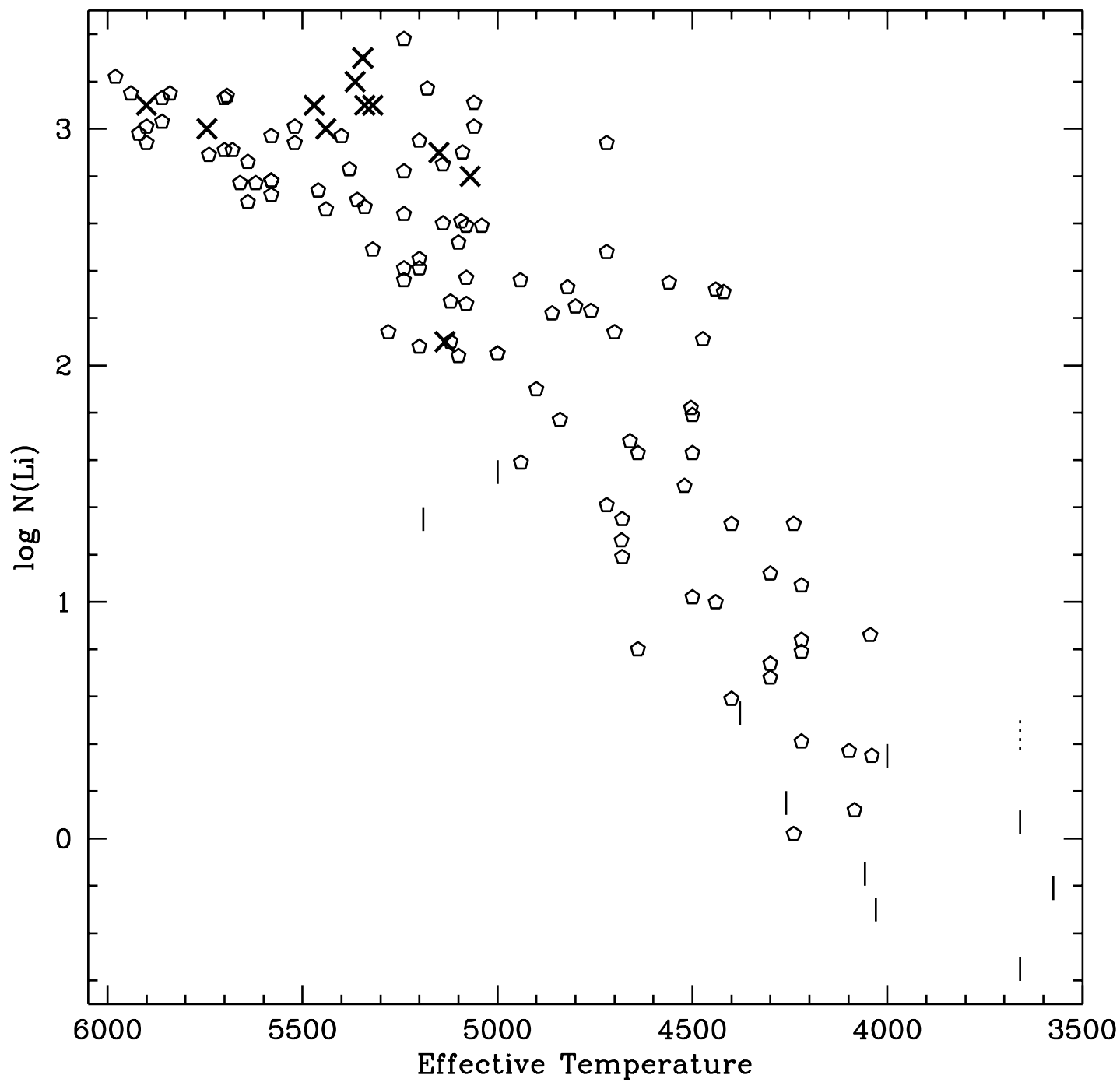
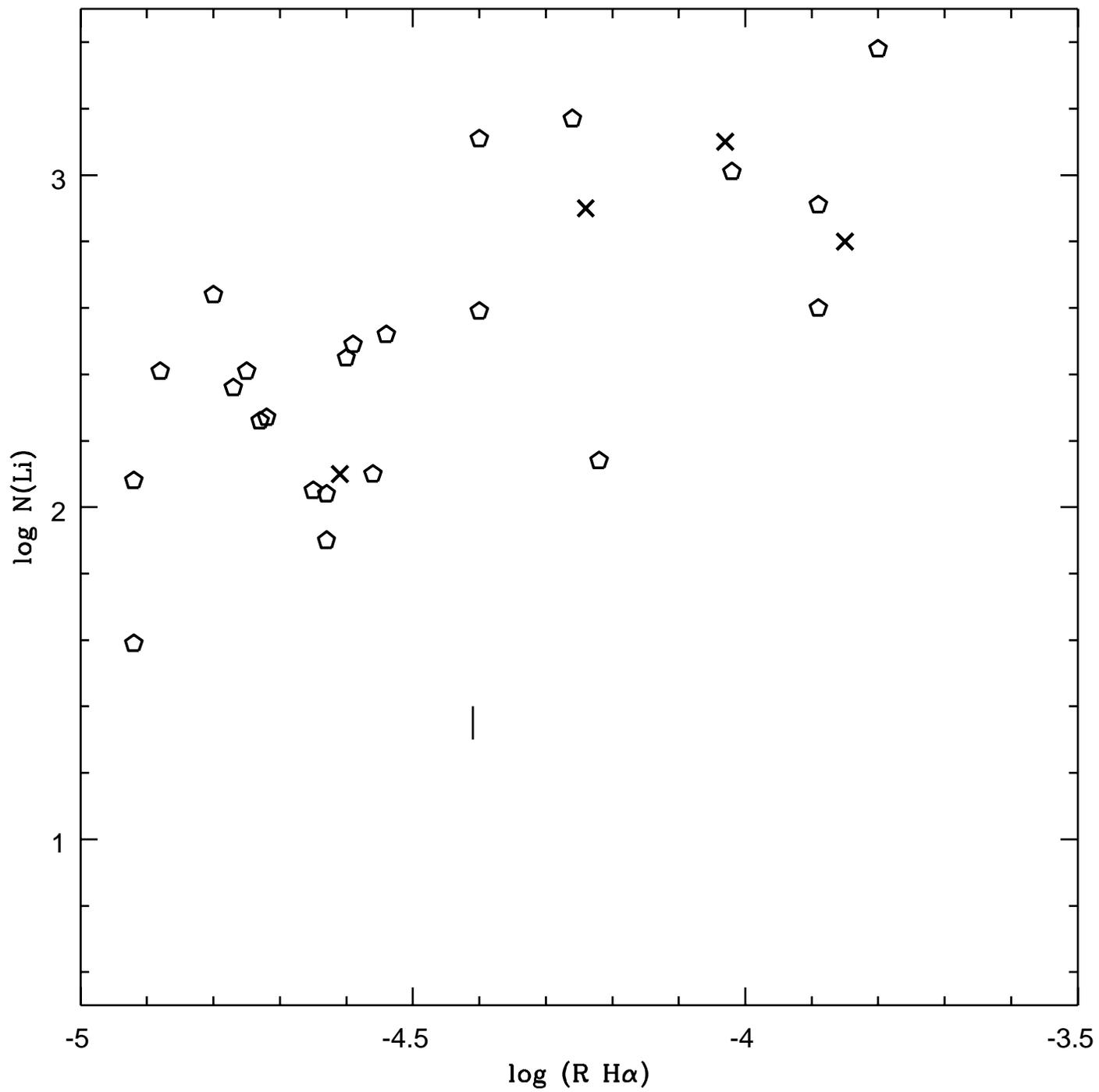
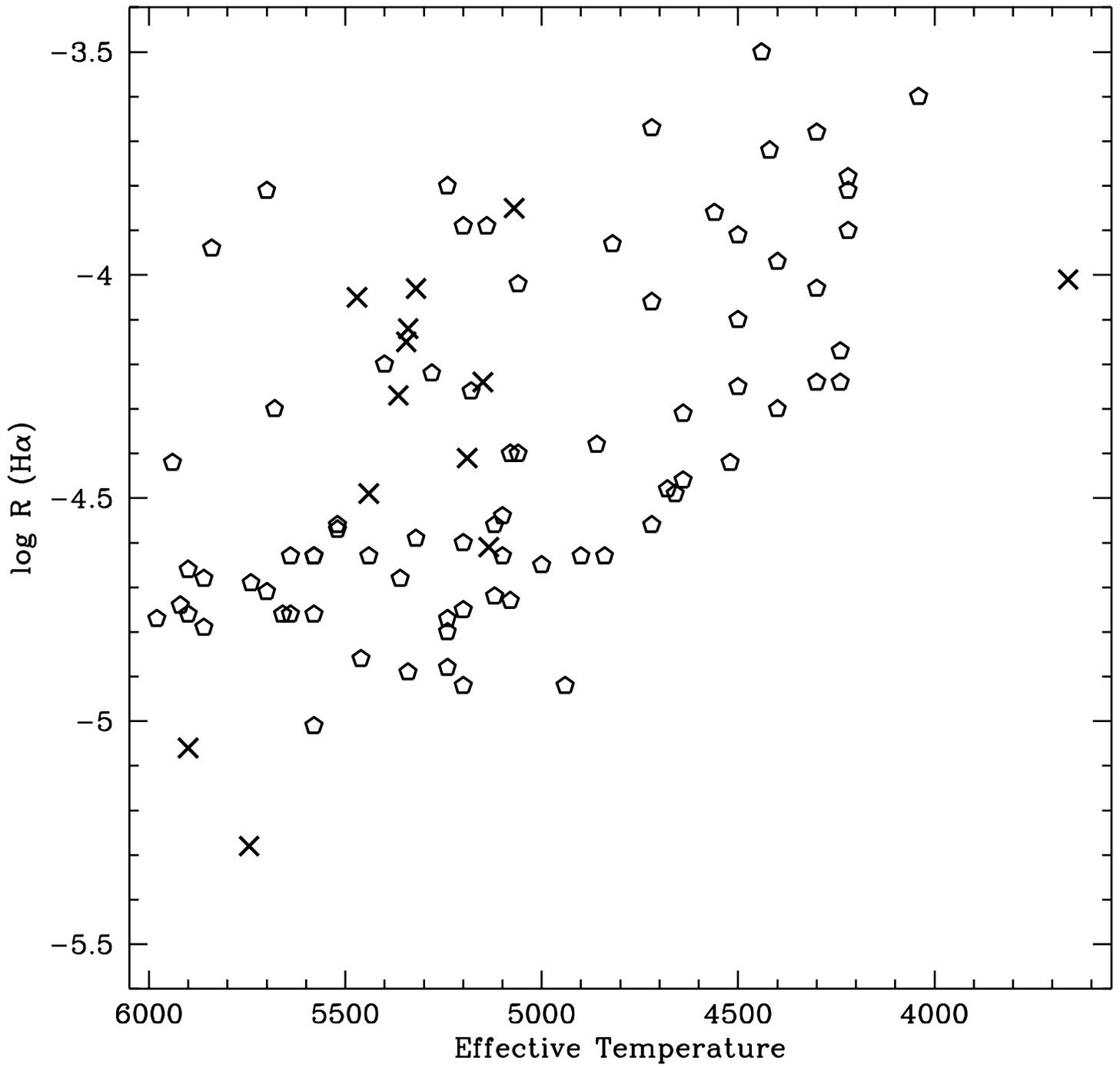
\end{document}